\documentclass[12pt]{iopart}

\usepackage{epsfig}
\usepackage{color}

\begin{document}  

\title[Measurement of $^{12}$C(e,e$^{\prime}$p)$^{11}$B at High Missing Momentum]
{Measurement of the $^{12}$C(e,e$^{\prime}$p)$^{11}$B Two-Body Breakup Reaction at High Missing Momentum}

\author{
P.~Monaghan$^{1,2}$, R.~Shneor$^3$, R.~Subedi$^4$, B.~D.~Anderson$^4$, K. Aniol$^5$,
J.~Annand$^6$, J.~Arrington$^7$, H.~Benaoum$^8$, F.~Benmokhtar$^{9,10}$,
P.~Bertin$^{11}$, W.~Bertozzi$^1$, W.~Boeglin$^{12}$, J.~P.~Chen$^{13}$,
Seonho~Choi$^{14}$, E.~Chudakov$^{13}$, C.~Ciofi~degli~Atti$^{15}$, E.~Cisbani$^{16}$,
W.~Cosyn$^{17}$, B.~Craver$^{18}$, C.~W.~de~Jager$^{13}$, R.J.~Feuerbach$^{13}$, E.~Folts$^{13}$,
S.~Frullani$^{16}$, F.~Garibaldi$^{16}$ O.~Gayou$^1$, S.~Gilad$^1$, R.~Gilman$^{9,13}$,
O.~Glamazdin$^{19}$, J.~Gomez$^{13}$, O.~Hansen$^{13}$, D.~W.~Higinbotham$^{13}$,
T.~Holmstrom$^{20}$, H.~Ibrahim$^{21,34}$, R.~Igarashi$^{22}$, E.~Jans$^{23}$,
X.~Jiang$^9$, L.~Kaufman$^{24}$, A.~Kelleher$^{20}$, 
A.~Kolarkar$^{25}$,
E.~Kuchina$^9$, G.~Kumbartzki$^9$, J.~J.~LeRose$^{13}$,
R.~Lindgren$^{18}$, N.~Liyanage$^{18}$, D.~J.~Margaziotis$^5$, P.~Markowitz$^{12}$,
S.~Marrone$^{16}$, M.~Mazouz$^{26}$, D.~Meekins$^{13}$, R.~Michaels$^{13}$,
B.~Moffit$^{13,20}$, H.~Morita$^{27}$, S.~Nanda$^{13}$, C.~F.~Perdrisat$^{20}$,
E.~Piasetzky$^3$, M.~Potokar$^{28}$, V.~Punjabi$^{30}$, Y.~Qiang$^1$,
J.~Reinhold$^{12}$, B.~Reitz$^{13}$, G.~Ron$^3$, G.~Rosner$^6$,
J.~Ryckebusch$^{17}$, A.~Saha$^{13}$, B.~Sawatzky$^{13,18,31}$,
J.~Segal$^{13}$, A.~Shahinyan$^{32}$, S.~\v{S}irca$^{28,29}$, K.~Slifer$^{18,31}$,
P.~Solvignon$^{13,31}$, V.~Sulkosky$^{1,20}$, N.~Thompson$^6$,
P.~E.~Ulmer$^{21}$, G.~M.~Urciuoli$^{16}$,
E.~Voutier$^{26}$, K.~Wang$^{18}$, J.~W.~Watson$^4$,
L.B.~Weinstein$^{21}$, B.~Wojtsekhowski$^{13}$, S.~Wood$^{13}$,
H.~Yao$^{31}$, X.~Zheng$^{1,7}$, and L.~Zhu$^{33}$
}

%
\address{$^1$ Massachusetts Institute of Technology, Cambridge, Massachusetts 02139, USA}
\address{$^2$ Hampton University, Hampton, Virginia, 23668, USA}
\address{$^3$ Tel Aviv  University, Tel Aviv 69978, Israel} 
\address{$^4$ Kent State University, Kent, Ohio 44242, USA}
\address{$^5$ California State University Los Angeles, Los Angeles, California 90032, USA}
\address{$^6$ University of Glasgow, Glasgow G12 8QQ, Scotland, UK}
\address{$^7$ Argonne National Laboratory, Argonne, Illinois, 60439, USA}
\address{$^8$ Syracuse University, Syracuse, New York 13244, USA}
\address{$^9$ Rutgers, The State University of New Jersey, Piscataway, New Jersey 08855, USA}
\address{$^{10}$ University of Maryland, College Park. Maryland 20742, USA}
\address{$^{11}$ Laboratoire de Physique Corpusculaire, F-63177 Aubi\`{e}re, France}
\address{$^{12}$ Florida International University, Miami, Florida 33199, USA}
\address{$^{13}$ Thomas Jefferson National Accelerator Facility, Newport News, Virginia 23606, USA}
\address{$^{14}$ Seoul National University, Seoul 151-747, Korea}
\address{$^{15}$ INFN, Sezione di Perugia , Via A. Pascoli I-06123,  Perugia, Italy}
\address{$^{16}$ INFN, Sezione Sanit\'{a} and Istituto Superiore di Sanit\'{a}, Laboratorio di Fisica, I-00161 Rome, Italy}
\address{$^{17}$ Ghent University, Proeftuinstraat 86, B-9000 Gent, Belgium}
\address{$^{18}$ University of Virginia, Charlottesville, Virginia 22904, USA}
\address{$^{19}$ Kharkov Institute of Physics and Technology, Kharkov 310108, Ukraine}
\address{$^{20}$ College of William and Mary, Williamsburg, Virginia 23187, USA}               
\address{$^{21}$ Old Dominion University, Norfolk, Virginia 23508, USA}
\address{$^{22}$ University of Saskatchewan, Saskatoon, Saskatchewan, Canada S7N 5E2}
\address{$^{23}$ Nationaal Instituut voor Kernfysica en Hoge-Energiefysica, Amsterdam, The Netherlands}
\address{$^{24}$ University of Massachusetts Amherst, Amherst, Massachusetts 01003, USA}
\address{$^{25}$ University of Kentucky, Lexington, Kentucky 40506, USA}
\address{$^{26}$ Laboratoire de Physique Subatomique et de Cosmologie, 38026 Grenoble, France}
\address{$^{27}$ Sapporo Gakuin University, Bunkyo-dai 11, Ebetsu 069, Hokkaido, Japan}
\address{$^{28}$ Institute ``Jo\v{z}ef Stefan'', 1000 Ljubljana, Slovenia}
\address{$^{29}$ Dept. of Physics, University of Ljubljana, 1000 Ljubljana, Slovenia}
\address{$^{30}$ Norfolk State University, Norfolk, Virginia 23504, USA}
\address{$^{31}$ Temple University, Philadelphia, Pennsylvania 19122, USA}
\address{$^{32}$ Yerevan Physics Institute, Yerevan 375036, Armenia}
\address{$^{33}$ University of Illinois at Urbana-Champaign, Urbana, Illinois 61801, USA}
\address{$^{34}$ Cairo University, Giza 12613, Egypt}

\begin{abstract}
The five-fold differential cross section for the
$^{12}$C$(e,e^{\prime}p)^{11}$B reaction was determined over a missing
momentum range of 200 -- 400~MeV/c, in a kinematics regime with $x_{B}
> 1$ and $Q^{2} = 2.0 \: \left({\rm GeV/c} \right) ^{2}$.  A
comparison of the results with previous lower missing momentum data
and with theoretical models are presented.  The extracted distorted
momentum distribution is shown to be consistent with previous data and
extends the range of available data up to 400~MeV/c.  The theoretical
calculations are from two very different approaches, one mean field
and the other short range correlated; yet for this system the two
approaches show striking agreement with the data and each other up to
a missing momentum value of 325~MeV/c.  For larger momenta, the
calculations diverge which is likely due to the factorization
approximation used in the short range approach.
\end{abstract}


\bibliographystyle{unsrt}


%

While the independent particle nuclear shell model has enjoyed much
success in predicting properties of nucleons in the nucleus up to the
Fermi momentum, of approximately 250~MeV/c, the model breaks down at
larger momenta~\cite{jjkelly} and fails as well for some observables
at very low momenta.  For example, the observed
spectroscopic strength, a multiplicative factor required to match the
predicted cross sections with data for valence orbital knockout,
averages around 0.65 instead of 1.0 as predicted by the independent
particle shell model. One explanation is that there are
nucleon-nucleon correlations present, which are neglected in
independent particle calculations. The effect of such correlations
would be to deplete valence states occupied below the
Fermi momentum and enhance continuum states occupied above
the Fermi momentum. In the lab, this translates into shifting strength
from low missing momentum ($\vec{p}_{m}$, the momentum of the
undetected residual system) and low missing energy, $E_{m}$ (which
accounts for the separation energy for removing a proton from the
target nucleus and any excitation of the residual system) to higher
missing momentum and energy in the A$(e,e^{\prime}p)$ reaction.

Many experiments and much work has been done to study nucleon-nucleon
correlations and, while a complete review is outside the scope of this
paper, ref.~\cite{Arrington:2011xs} provides a clear outline and
discussion of the experimental evidence for suitable kinematics for
studying nucleon-nucleon correlations. In particular, with a
four-momentum transfer squared, $Q^{2} = -q_{\mu}q^{\mu} =
|\vec{q}^{2}| - \omega^{2} > 1$ [GeV/$c$]$^2$ and Bjorken scaling
variable $1 < x _{B} = Q^{2}/2m_{p}\omega < 2$, where $\omega$ and
$\vec{q}$ are the energy and three-momentum transfer respectively from
the electron are preferable for studying high $p_{m}$ nucleon-nucleon
correlations. Such kinematics minimize competing effects such as
meson-exchange currents (MEC), isobar configurations (IC) and final
state interactions (FSI) which can mask the correlation.

In this work we examine the $^{12}$C$(e,e^{\prime}p)^{11}$B two-body
breakup channel; parallel studies of short-range correlations related
to the multi-nucleon knockout reaction channels
$^{12}$C$(e,e^{\prime}pp)$ and $^{12}$C$(e,e^{\prime}pn)$ have been
published separately~\cite{ranprl,subedi}. The experiment was
performed in Hall A at the Thomas Jefferson National Accelerator
Facility (JLab), using the High Resolution Spectrometers
(HRS)~\cite{Alcorn:2004sb}. The data were taken at a fixed electron
beam energy of 4.627~GeV incident on a 0.25 mm thick natural-carbon
foil target. The scattered electrons were detected in the left HRS at
a central scattering angle and momentum of $19.5^{\circ}$ and
3.762~GeV/$c$, respectively. This fixed the electron kinematics,
resulting in a central three-momentum transfer of $|\vec{q}| = 1.66$
GeV/c and energy transfer of $\omega = 0.865$~GeV, which corresponds
to a four momentum transfer squared of $Q^{2} \simeq 2$ (GeV/c)$^{2}$
and Bjorken scaling variable, $x_{B} \simeq 1.23$. The knocked-out
protons were detected in the right HRS at an angle of $\theta_{p} =
40.1^{\circ}$ and a central proton momentum of $|\vec{p}_{p}| =
1.45$~GeV/c.  This spectrometer setting provided a continuous coverage
of missing momentum $\vec{p}_{m} = \vec{q} - \vec{p}_{p}$ from 200 to
400~MeV/$c$ for the $^{12}$C$(e,e^{\prime}p)^{11}$B reaction.

%
%

\begin{figure}[ht]
\centerline{\epsfig{width=\linewidth,file=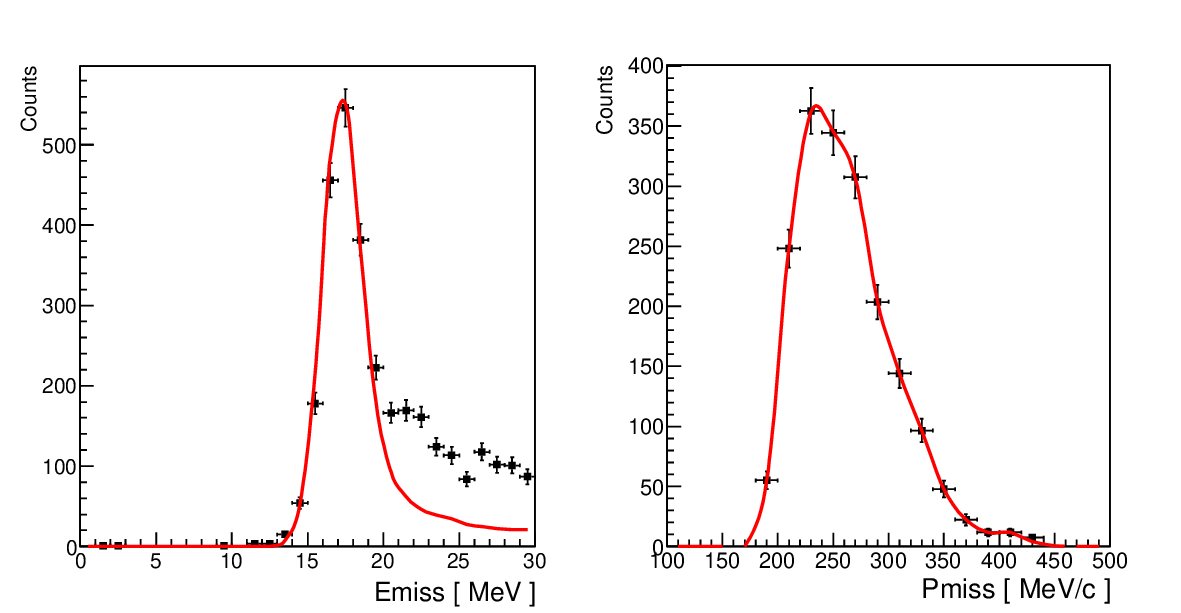}}
\caption{LEFT: Number of counts versus missing energy for the proton
   spectrometer central momentum of 1.45~GeV/c. The peak corresponds to
   proton knockout from the lowest states of carbon, predominantly the
   p$_{3/2}$ state. The red curve shows the simulated data set used for
   this analysis of the p$_{3/2}$-shell knockout and illustrates the
   good agreement between the data and the simulation. The simulation
   curve also shows the tail of events extending out to large missing
   energy values, which result from real photons radiated out by the
   electron.  RIGHT: The resulting missing momentum distribution after
   applying a cut on missing energy of 20 MeV to select scattering
   events to the $^{11}$B ground state. The missing momentum data were
   fit by varying the cross section model in the simulation. The red
   curve shows the final agreement between the data and the
   simulation.}
\label{kinfig}
\end{figure}

The peak in the missing energy distribution shown in
Figure~\ref{kinfig} (left-hand plot) results predominantly from
knockout of protons from the p$_{3/2}$ shell, leaving the residual
$^{11}$B nucleus in its ground state. There is also a {\it shoulder}
of events which is observed around 22~MeV in Figure~\ref{kinfig}; this
is due to the $^{11}$B nucleus being left in one of several low-lying
excited states. Since the missing energy resolution of the experiment
($\sim$3~MeV) was insufficient to separate the individual states of
$^{11}$B, two simulations were prepared in order to determine correct
location of the cut in missing energy which was applied to separate
the ground state and excited state contributions. One simulation was
fit to the ground state peak and the second simulation was made for
the low-lying excited states. By summing the resultant yield from the
two simulations and fitting it to the data, the contribution of only
the ground state into the {\it shoulder} region could be separated
out. This allowed the application of the missing energy cut at 20~MeV
to optimize only the ground state simulation with minimal
contamination from the low-lying excited states. In
Figure~\ref{kinfig} (left-hand plot) the red curve shows the results
of a simulation normalized to the data that includes radiation of real
photons by the electron (the `radiative tail'). There is a large
number of events at missing energies greater than 20~MeV resulting
from knockout of protons from the s-shell in addition to knockout of
protons from correlated nucleon--nucleon pairs.

The $(e,e^{\prime}p)$ events were selected by placing a 1.1~ns cut
around the coincidence timing peak as well as using the HRS pion
rejector to suppress the small amount of pion background.  The
resulting event sample, contained less then 1\% random events.  The
only other cuts on the data were the nominal HRS phase space
cuts~\cite{Alcorn:2004sb}, on momentum, $| dp/p | <0.04$ and
angular cuts of $| \theta | < 0.05$ radians and $|
\phi | < 0.03$ radians about the spectrometers central ray. These
cuts discarded events from the edges of the spectrometer acceptances.

A full simulation, including energy losses, multiple scattering,
internal and external radiation and spectrometer resolutions was
performed. The same set of acceptance cuts, which was applied to the
data, was also applied to the simulation. The simulation program MCEEP
({\bf M}onte {\bf C}arlo for {\bf e}, {\bf e$^{\prime}$
p})~\cite{MCEEP} was used to extract the five-fold differential cross
section from the data by using an iterative procedure to adjust the
radiated $^{12}$C$(e,e^{\prime}p)$ cross section in the simulation
until the simulated yield agreed with the experimental yield in each
missing momentum bin. The right-hand plot of Figure~\ref{kinfig} shows
the final agreement between the simulation (red curve) and the missing
momentum data after selecting the ground state data.
                                                  
The cross section model used in this analysis was based on the
factorized distorted wave impulse approximation (DWIA) and is defined
as~\cite{deforest}
\begin{equation}
  \frac{d^{6} \sigma}{d \Omega_{e} d \Omega_{p} dE_{e} dE_{p}} = E_{p}
  \: p_{p} \: \sigma_{cc2} \: S_{D}(\vec{p}_{m},E_{m}),
\end{equation}

\noindent where $\sigma_{cc2}$ is the single-nucleon off-shell
cross section prescription of de Forest~\cite{deforest}.  The
$\sigma_{cc2}$ prescription is a current-conserving off-shell
extrapolation of the on-shell current obtained from the Dirac
equation. This prescription includes explicitly the four-momentum
transfer ($q^{\mu}$) in the nucleon current calculation, whereas the
$\sigma_{cc1}$ prescription does not; further details are given
in~\cite{deforest}. $S_{D}(\vec{p}_{m},E_{m})$ is the proton part of
the distorted spectral function and is the probability of detecting a
proton in the nucleus with momentum $\vec{p}_{m}$ and energy
$E_{m}$. 

Integrating Eq. (1) over the missing energy peak in the discrete part
of the $^{11}$B spectrum leads to the five-fold differential
cross section for a specific state
\begin{equation}
  \label{xseqn}
  \frac{d^{5} \sigma}{d \Omega_{e} d \Omega_{p} dE_{e}} = K
  \sigma_{cc2} \: \int_{\Delta E_{m}} S_{D} \left( \vec{p}_{m},E_{m'}
  \right) dE_{m'},
\end{equation}
\noindent where $K = E_{p} \: p_{p} \: \eta^{-1}$ and $\eta$ is the
recoil factor for scattering to a bound state. When integrating
Eq. (1) over the missing energy peak to obtain the five-fold
differential cross section, $eta$ is the Jacobian which arises and
given by
\begin{equation}
  \eta = 1 - \frac{E_{p}}{E_{r}} \frac{\vec{p_{p}} \cdot \vec{p_{r}}}{
  |\vec{p_{p}}|^{2} },
\end{equation}
where $E_{r}$ and $\vec{p}_{r}$ are the energy and momentum of the
recoiling system. The integral is performed over $\Delta E_{m}$ which
is the range of missing energy for the specific state being analyzed
to account for the natural width of the final state along with the
experimental energy resolution.  In this analysis, a range of
integration of 0 to 20 MeV was used to select events from the
p$_{3/2}$-shell.

At sufficiently low values of the missing energy, the spectral
function is centered around specific values of $E_{m}$ and it can be
assumed to factorize into two functions,
\begin{equation}
  S_{D} \left( \vec{p}_{m}, E_{m} \right) = \sum_{\alpha}
  n_{\alpha} \left( \vec{p}_{m} \right)
  f_{\alpha} \left( E_{m}^{\alpha} \right) \;,
\end{equation}

\noindent where $f_{\alpha} \left( E_{m}^{\alpha} \right)$ is the
missing energy distribution for state $\alpha$ and is sharply peaked
about $E_{m}^{\alpha}$ and $n_{\alpha} \left( \vec{p}_{m} \right)$ is
the missing momentum distribution for state $\alpha$. The above
considerations allow the cross section to be written as,
\begin{equation}
  \frac{d^{5} \sigma}{d \Omega_{e} d \Omega_{p} dE_{e}} = 
  K \sigma_{cc2} \: n_{\alpha} \left( \vec{p}_{m} \right)
  \int f_{\alpha} \left( E_{m} \right) dE_{m} .
  \label{sigma5}
\end{equation}

\noindent Since the missing mass distribution function for a bound
state is a delta function, the model cross section can
be modified by adjusting the input momentum distribution
$n_{\alpha}(\vec{p}_{m})$ in the simulation to agree with the
measured cross sections.

An iterative procedure was used to fit the simulated yield to the
experimental data. Starting with an initial input momentum function,
the simulation was run and the resulting simulated yield as a function
of missing momentum compared to the counts in the experimental data,
with the same set of cuts being applied to both the simulation and the
data. The difference between the simulated yield and the experimental
data was used to then modify the input momentum function for the next
simulation run.  This iteration procedure continued until the
simulated yield agreed with the experimental data to one percent.

The dominant systematic error in this analysis arose from the
selection of scattering events from the $^{12}$C p$_{3/2}$-shell to
the $^{11}$B ground state.  This dependence was evaluated by shifting
the limit on the $E_{m}$ by $\pm$1 MeV around 20~MeV and refitting the
missing momentum distribution.  This yielded a cut dependence of 5.9\%
in the extracted cross section.  Varying other cuts showed no
significant change in the resulting cross sections.  The next leading
source of systematic error was the normalization of the luminosity by
comparison with the cross section from hydrogen elastic
scattering. For this, the scattering data from a 4cm extended liquid
hydrogen target were compared to MCEEP simulations which used a form
factor model derived from a fit to world data~\cite{arrington:2003qk},
this resulted in a 4.5\% uncertainty\footnote{This uncertainty is
conservative because the hydrogen data were taken using an extended
target cell, whereas the physics data were taken on a carbon foil
(point) target. Also, a proper background subtraction for the target
cell could not be done as the experiment did not take any data with a
dummy target cell}.  Other minor sources of uncertainty included an
absolute tracking efficiency uncertainty of 1.1\%, an uncertainty on
the radiative corrections of 1.0\% and the uncertainty from the
acceptance cuts of $<$ 1.0\%.  The combined total systematic
uncertainty was 7.6\%.

\begin{figure}[ht]
\centerline{\epsfig{width=\linewidth,file=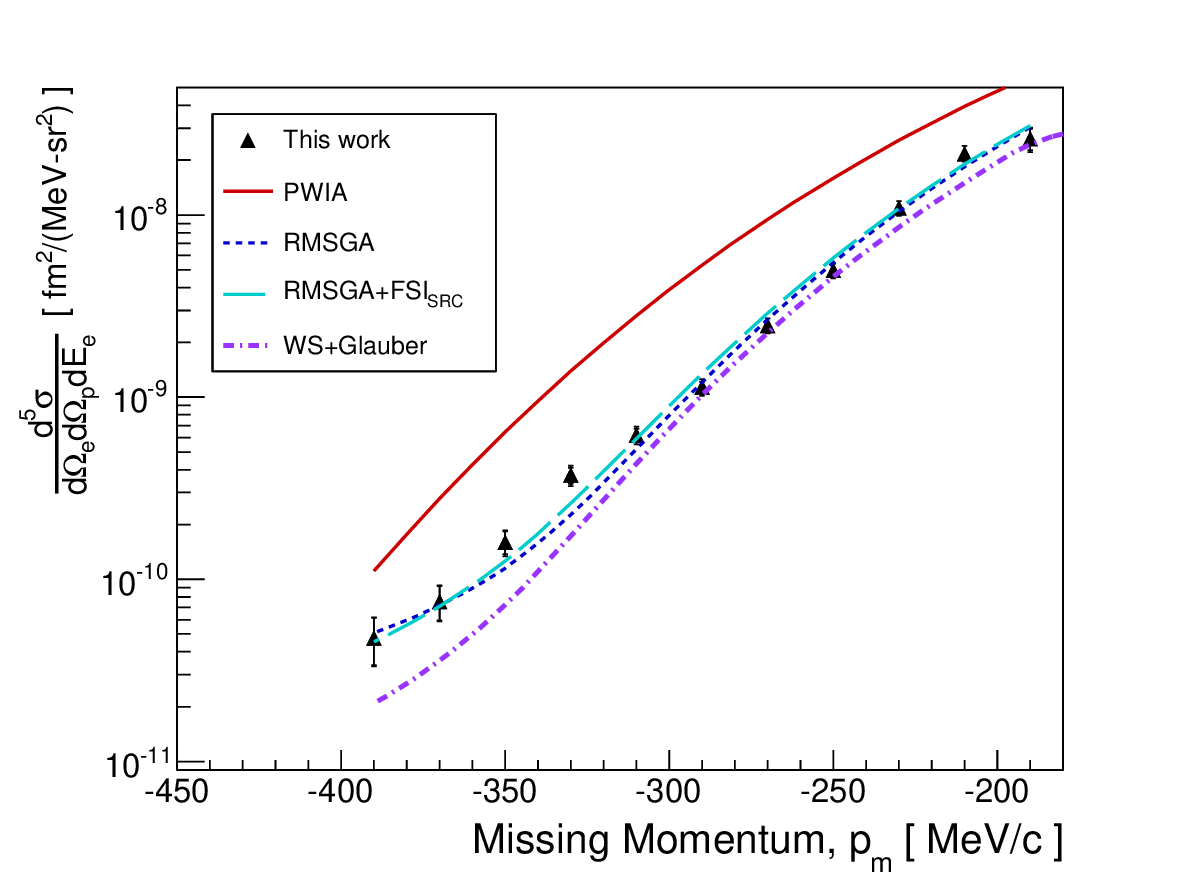}}
\caption{Experimental five-fold differential cross section extracted
  from the data are compared to several different theoretical
  calculations.  The ground-state wave function for the WS+Glauber
  calculation includes the effects of correlations, while the one used
  in the RMSGA calculations does not; no other normalization
  (spectroscopic factor) has been applied to the calculations.  Each
  data point is plotted at the center of a 20 MeV wide missing
  momentum bin. The difference between the PWIA calculation and the
  others demonstrates the importance of properly including final-state
  interactions to describe the experimental results.}
\label{xsfig}
\end{figure}

The resulting five-fold differential cross sections as a function of
missing momentum is shown in Fig.~\ref{xsfig}. The cross section data
spans a range of missing momentum of $-200$ to $-400$~MeV/c. Although the
electron spectrometer had a fixed momentum and angle setting
throughout the experiment, the fact that both spectrometers have
finite acceptances leads to each data point having slightly different
values of $Q^{2}$ and $\omega$ for each missing momentum bin. 

Also shown in Fig.~\ref{xsfig} are four curves produced by
calculations utilizing different models. Each calculation has been
performed using the appropriate kinematics for each missing momentum
bin, rather than only the central spectrometer setting.
The RMSGA and RMSGA + FSI$_{SRC}$ curves in Fig.~\ref{xsfig} are
unfactorized relativistic formulations calculations by W.~Cosyn and
J.~Ryckebusch~\cite{Ryckebusch2003226}.  The bound-state wave
functions are solutions to the Dirac equation with scalar and vector
potentials fitted to ground-state nuclear properties. The final state
interactions are modelled on rescattering of a fast proton from a
composite target containing $A-1$ {\it frozen} spectator nucleons. The
curve labelled `RMSGA+FSI$_{SRC}$' differs from the `RMSGA'
calculation in that it has been extended to include short-range
correlation effects in the final state
interactions~\cite{ryckebuschPhysRevC.77.034602}. These correlations
create local fluctuations in the nuclear density, modifying the
attenuation of the scattered proton in the nuclear medium.

The WS+Glauber curve in Fig.~\ref{xsfig} corresponds to a factorized
calculation by M. Alvioli, C.~Ciofi~degli~Atti and
H.~Morita~\cite{claudio}.  This model uses many-body variational
correlated wavefunctions resulting from a cluster expansion solution
of the non-relativistic Schroedinger equation with realistic
nucleon-nucleon interactions and Woods-Saxon (WS) single particle
wavefunctions~\cite{claudio}.  The final-state interactions are
modeled using an improved Glauber approach~\cite{glauber} to describe
the rescattering of the knocked-out
proton~\cite{Morita:1999gm,CiofidegliAtti:2006jc,CiofidelgiAtti:2007qu}. This
calculation includes ground state correlations in the initial wave
function which result in a reduction of the cross section by a factor
of 0.8. This has the effect of reducing the occupation number of the
$1p_{3/2}$ shell predicted by an independent particle shell model.

Plane Wave Impulse Approximation (PWIA) calculations were done
independently by both groups to ensure the same baseline calculation
and gave the same result which is why only one PWIA curve is shown in
Fig. 2.  More striking is the agreement between the WS+Glauber
calculation and full the RMSGA calculation.  Only at the highest
momenta does one start to see a deviation between the curves and
preference of the data to the RMSGA calculation.  This is likely due
to the fact that at this time the WS+Glauber calculation is factorized
while the RMSGA is unfactorized.

The experimental distorted momentum distribution is extracted
from the cross section data by dividing out the kinematic factor and
the single nucleon off-shell cross section terms, $\sigma_{cc2}$, in
Eq.~\ref{xseqn}. This was accomplished by running another MCEEP
simulation, with all of the input parameters unchanged, except now
with a uniform input momentum function. This meant that all of the
same averaging over the same missing momentum bins as was done for the
cross section extraction was repeated for this simulation to generate
just the kinematic factor and the single nucleon off-shell cross
section. The distorted momentum distribution is then given by,
\begin{equation}
  n_{distorted}(p_{m}) = \left\langle \frac{d^{5} \sigma}{d \Omega_{e}
  d \Omega_{p} dE_{e}} \right\rangle_{exp} / \left\langle K
  \sigma_{cc2} \right\rangle_{unit}.
\end{equation}

\begin{figure}[ht]
\centerline{\epsfig{width=\linewidth,file=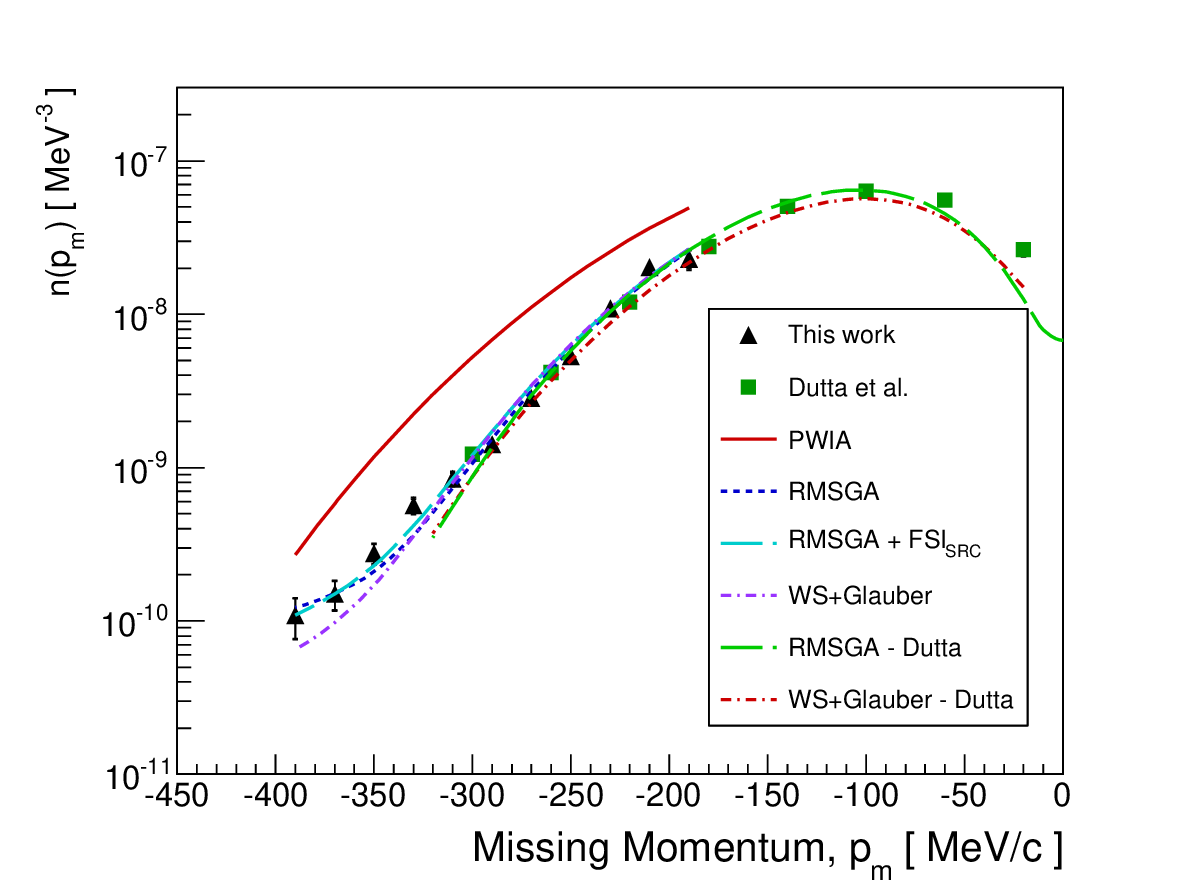}}
\caption{Comparison of experimental distorted momentum distribution
  extracted from two different experiments with theoretical
  calculations based on the RMSGA and WS+Glauber approaches, as well
  as a simple PWIA calculation. The agreement with the experimental
  data shows the same general trends as observed in the cross section
  comparison and shows in general this reaction can be well described
  with either a mean field or WS+Glauber calculation over nearly
  the entire momentum range. Note, the curves labelled
  {\it `Dutta'} were made using the kinematics from the Hall C
  experiment~\cite{dutta68}, which are slightly different from our
  kinematics. The error bars for both sets of data are plotted,
  although the logarithmic scale makes their observation difficult for
  all but the largest errors.}
\label{momfig}
\end{figure}

The experimental distorted momentum distribution from this experiment
covers a range in missing momentum of $-200$ to $-400$~MeV/c, which
overlaps with data from a previous experiment in Hall C at
JLab~\cite{dutta68}.  A comparison of the experimental distorted
momentum distributions from both experiments along with calculated
momentum distributions arising from the models used to calculate the
cross sections in Fig.~\ref{xsfig} are shown in Fig.~\ref{momfig}.
The data from the Hall C experiment shown here were taken at $Q^{2} =
1.8$~(GeV/c)$^{2}$ and include a cut on missing energy of $15 < E_{m}
< 25$~MeV to select the $p$-shell; however, this missing energy range
does include some contribution from $s$-shell knockout. The Hall C
analysis includes a factor of $(2j+1)$ for the multiplicity of the
shell being considered.  This analysis extracted the cross section per
nucleon and so to make this comparison, our data are multiplied by a
factor of 4. 

The RMSGA and WS+Glauber calculations were performed for the
applicable kinematics from both experiments for the data shown and
hence there are two different sets of curves in Fig.~\ref{momfig}.  As
Fig.~\ref{momfig} shows, the two experiments agree in the overlap
momentum region from $-200$ to $-300$~MeV/c. The data from this
experiment extends the experimental momentum distribution to
$-400$~MeV/c. The resulting momentum distributions from the PWIA,
RMSGA and WS+Glauber calculations are compared to the data in
Fig.~\ref{momfig}. The RMSGA and WS+Glauber calculations are in good
agreement with the data from both experiments and each other up to a
missing momentum around 325~MeV/c, where they start to diverge
slightly likely due to the factorization approximation in the
WS+Glauber calculation.

Since our data is predominantly from the p$_{3/2}$ shell, the
occupation number for that shell in the carbon ground state can be
inferred from the theoretical calculation. The number of protons in a
sub-shell of the ground state is given by the integral of the momentum
distribution of that shell. For our data, since we only have
experimental data covering the missing momentum range of $-200$ to
$-400$\~MeV/c, and not the full momentum range, the integral has to be
performed using the fact that we have good agreement with the
theoretical models. For this result, the occupation number was
determined by integral of the WS+Glauber calculation over a range of
missing momentum from $-20$ to $-400$~MeV/c. This integral is
precisely 3.2, but is the total occupation number for the
p-shell. This occupation number arises from 4$\times$0.8, corresponding to
the factor of 0.8 which was already included in the initial state
wavefunction to account for short-range correlations. This does not
tell us anything about separate occupation probabilites of either the
p$_{3/2}$ or p$_{1/2}$ shells in carbon as these are also governed by
long-range correlations. However, shell model calculations by other
groups have shown that $\sim 70\%$ of p-shell protons occupy the
p$_{3/2}$ ground state. 
This means the integral value is further reduced by a factor of 0.7, resulting
in an occupation number of 2.24 for the p$_{3/2}$ shell.

For completeness, the experimental cross section and extracted
momentum distribution results are shown in Table~\ref{grdtable}. The
statistical uncertainties for each data point are quoted, with the
same statistical uncertainty applicable to both the cross section and
the momentum distribution. The systematic uncertainty, includes
normalization, kinematic and event selection uncertainties which were
all added together in quadrature, to produce a global value of
7.6\%. The quoted cross section numbers are given by the
experimentally-normalised Monte Carlo simulation at the quoted central
kinematic values for each bin. The average values of $Q^{2}$ and
$\omega$ as well as their RMS widths for each missing momentum bin are
also provided.

\begin{table}
 \begin{center}
  \begin{tabular}{lllll}  \hline
$p_{m}$ 	& $\frac{d^{5}\sigma}{d\Omega_{e}d\Omega_{p}dE_{e}} \pm
    \delta_{stat}$ & $n(p_{m})$ & $\bar{Q^{2}} \pm \sigma_{Q^{2}}$ & $\bar{\omega} \pm \sigma_{\omega}$ \\ 
 $[$MeV/c]  	& [fm$^{2}$/(MeV-sr$^{2}$)] & [MeV$^{-3}$] & [(GeV/c)$^{2}$] & [MeV] \\ \hline 
190 		& 2.62e-08 $\pm$ 13.5\%  & 2.29e-08  & 1.70 $\pm$ 0.018 & 852 $\pm$ 4.2 \\
210 		& 2.18e-08 $\pm$ 6.3\%   & 2.03e-08  & 1.72 $\pm$ 0.033 & 846 $\pm$ 8.4 \\
230 		& 1.09e-08 $\pm$ 5.3\%   & 1.09e-08 & 1.74 $\pm$ 0.046 & 841 $\pm$ 12.5 \\
250 		& 4.95e-09 $\pm$ 5.4\%   & 5.31e-09 & 1.77 $\pm$ 0.060 & 836 $\pm$ 16.6 \\
270 		& 2.47e-09 $\pm$ 5.7\%   & 2.85e-09 & 1.80 $\pm$ 0.075 & 830 $\pm$ 20.7 \\
290 		& 1.13e-09 $\pm$ 7.0\%   & 1.42e-09 & 1.83 $\pm$ 0.088 & 824 $\pm$ 23.9 \\
310 		& 6.20e-10 $\pm$ 8.3\%   & 8.48e-10 & 1.87 $\pm$ 0.103 & 819 $\pm$ 26.0 \\
330 		& 3.73e-10 $\pm$ 10.2\%  & 5.66e-10 & 1.92 $\pm$ 0.117 & 815 $\pm$ 27.1 \\
350 		& 1.60e-10 $\pm$ 14.5\%  & 2.76e-10 & 1.98 $\pm$ 0.128 & 814 $\pm$ 27.8 \\
370 		& 7.56e-11 $\pm$ 21.2\%  & 1.50e-10 & 2.05 $\pm$ 0.134 & 813 $\pm$ 28.0 \\
390 		& 4.77e-11 $\pm$ 29.3\%  & 1.09e-10 & 2.11 $\pm$ 0.130 & 813 $\pm$ 27.7 \\
  \hline
  \end{tabular}
   \caption{Results for the $^{12}$C(e,e$^{\prime}$p)$^{11}$B reaction
   data from the highest proton momentum spectrometer setting.  The
   statistical error quoted for the cross section is the same for the
   extraced momentum distribution. The cross section numbers are given
   by the experimentally-normalised Monte Carlo simulation at the
   central value of the kinematics quoted for each missing momentum
   bin. The central values of $Q^{2}$ and $\omega$ and their
   corresponding RMS widths $\sigma_{Q^2}$ and $\sigma_{\omega}$ are
   given for each missing momentum bin.  Not shown is the global
   systematic error of 7.6\%.}
   \label{grdtable}
 \end{center}
\end{table}

In summary, the experimental five-fold differential cross section for
the $^{12}$C$(e,e^{\prime}p)^{11}$B reaction has been extracted in a
previously unexplored kinematic region. The data extends over a range
of missing momenta from $-200$ to $-400$~MeV/c.  A comparison of our
data with calculations which include final state interactions by two
different approaches demonstrates the failure of PWIA while
highlighting the ability of modern mean field and short range
correlated calculations to both discribe the $^{12}$C(e,e'p)$^{11}B$
reaction over a large range of the missing momentum.  The variations
of the WS+Glauber calculations from the data at missing momenta above
325~MeV/c likely is the result of the factorization approximation in
these calculations.

The experimental distorted momentum distribution was also extracted
from the cross section data and compared with a previous experiment in
Hall C at JLab. The data from both experiments are consistent for the
region of missing momentum where they overlap. The theoretical
calculations also show good agreement with the data up to a missing
momentum around 325~MeV/c, where they diverge slightly.  In general
the agreement of the data with both the short range correlated
and mean field approaches is very good. Using the agreement of our
data with the calculations, the occupation number of the p$_{3/2}$
shell was inferred and found to be 2.24.


We would like to acknowledge the contribution of the Hall A
collaboration, the Hall A technical staff and the accelerator
operations staff. This work was supported by the
Israel Science Foundation, the US-Israeli Bi-national Scientific
Foundation, the UK Engineering and Physical Sciences Research Council,
the U.S. National Science Foundation, the U.S. Department of Energy
grants DE-FG02-94ER40844, DE-AC02-06CH11357, DE-FG02-94ER40818,
and U.S. DOE Contract No. DE-AC05-84150,
Modification No. M175, under which the Southeastern Universities
Research Association, Inc. operates the Thomas Jefferson National
Accelerator Facility. \\

\section*{References}


\end{document}